\documentclass[twocolumn,superscriptaddress,showpacs,aps,pra,%
longbibliography]{revtex4-1}

\usepackage{epsf}
\usepackage{amsmath}
\usepackage{amsfonts}
\usepackage{amssymb}
\usepackage{bm}
\usepackage{bbm}
\usepackage{nicefrac}
\usepackage{color}
\usepackage{pifont}
\usepackage{graphicx}
\usepackage{enumerate}
\usepackage{dcolumn}
\usepackage{maybemath}

\newcommand{\stdrule}{\rule[-1mm]{0mm}{4mm}}

\newcommand{\re}{\mathrm{Re}}
\newcommand{\im}{\mathrm{Im}}

\def\dd{{\mathrm{d}}}
\def\ii{{\mathrm{i}}}
\def\ee{{\mathrm{e}}}

\def\calZ{{\mathcal{Z}}}

\def\CaF2{{CaF$_2$}}

\def\GG{G}
\def\pp{d}
\def\ppp{p}

\def\tfrac#1#2{ {\textstyle{\frac{#1}{#2}} } }

\begin{document}

\title{Theory of noncontact friction for atom-surface interactions}

\author{U. D. Jentschura}

\affiliation{Department of Physics, Missouri University of Science
and Technology, Rolla, Missouri 65409, USA}

\author{M. Janke}
\author{M. DeKieviet}

\affiliation{Physikalisches Institut, 
Universit\"{a}t Heidelberg, INF226, 69120 Heidelberg, Germany}

\begin{abstract}
The noncontact (van der Waals) friction is an interesting physical effect 
which has been the subject of controversial scientific discussion.
The ``direct'' friction term
due to the thermal fluctuations of the electromagnetic field
leads to a friction force proportional to 
$1/\calZ^5$ (where $\calZ$ is the atom-wall 
distance). The ``backaction'' friction term 
takes into account the feedback of thermal fluctuations 
of the atomic dipole moment onto the motion of the 
atom and scales as $1/\calZ^8$.
We investigate noncontact friction effects for 
the interactions of hydrogen, ground-state helium
and metastable helium atoms with $\alpha$-quartz (SiO$_2$),
gold (Au) and calcium difluorite (CaF$_2$).
We find that the backaction term dominates over the 
direct term induced by the thermal electromagnetic fluctuations inside the 
material, over wide distance ranges. The friction coefficients obtained for gold 
are smaller than those for SiO$_2$ and CaF$_2$ by several orders of magnitude.
\end{abstract}


\maketitle


%
%
\section{Introduction}
\label{sec1}

Noncontact friction arises in atom-surface interactions;
the theoretical treatment has given rise to some
discussion~\cite{Le1989,Po1990,HoBr1992,*HoBr1993,Mk1995,ToWi1997,PeZh1998,%
*DeKy1999,*DeKy2001,*DeKy2002,DeKy2002review,%
VoPe1999,*VoPe2001prb,VoPe2002,VoPe2003,*VoPe2005,*VoPe2006,*VoPe2007,*VoPe2008,%
DoFuGoJe2001}. In a
simplified understanding, for an ion flying by a dielectric surface (``wall''),
the quantum friction effect can be understood in terms of Ohmic heating of the
material by the motion of the image charge inside the medium. Alternatively,
one can understand it in terms of the thermal fluctuations of the electric
fields in the vicinity of the dielectric, and the backreaction onto the motion
of the ion or atom in the vicinity of the ``wall''. 

It has recently been argued that one cannot separate the van-der-Waals force,
at finite temperature, from the friction effect~\cite{VoPe2002}.  The
backaction effect is due to the fluctuations of the atomic dipole
moment~\cite{VoPe2002},
which are mirrored by the wall and 
react back  onto the atom; this  leads to an additional contribution to the
friction force.  In contrast to the ``direct'' term created by the
electromagnetic field fluctuations inside the medium~\cite{ToWi1997}
(proportional to $1/\calZ^5$ where $\calZ$ is the atom-wall distance), the
backaction term leads to a $1/\calZ^8$ effect.  A comparison of the magnitude
of these two effects, for realistic dielectric response functions of materials,
and using a detailed model of the atomic polarizability, is the subject of the
current paper. While the $1/\calZ^8$ effect is parametrically suppressed
for large atom-wall separations, the numerical coefficients may still 
change the hierarchy of the effects.

We should also note that the direct term~\cite{ToWi1997,VoPe2002} can be
formulated as an integral over the imaginary part of the polarizability, and of
the dielectric response function of the material.
Recently, we found a conceptually
interesting ``one-loop'' dominance for the  imaginary part of the
polarizability~\cite{JeLaDKPa2015,JePa2015epjd1}. 
The imaginary part of the polarizability 
describes a process where the atom emits radiation at the 
same frequency as the incident laser radiation,
but in a different direction.
Note that, by contrast, Rabi flopping involves continuous absorption and 
emission into the laser mode; the laser-dressed states~\cite{ScZu1997,JeKe2004aop}
are superpositions of states $| g, n_L+1\rangle$ and 
$|e, n_L \rangle$, where $n_L$ is the number of laser 
photons while $|g\rangle$ and $|e\rangle$ denote the 
atomic ground and excited states.
{\em A priori}, this Rabi flopping may proceed off resonance.

\begin{figure}[thb]
\begin{center}
\begin{minipage}{1.0\linewidth}
\includegraphics[width=0.62\linewidth]{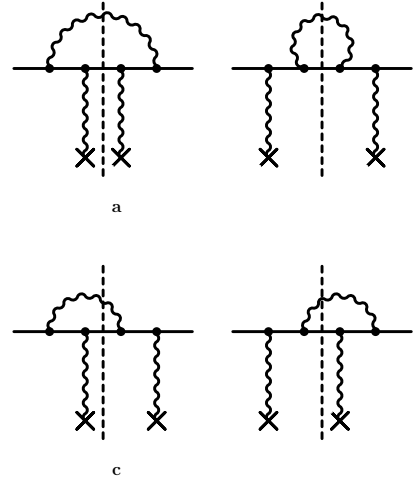}
\caption{\label{fig1} Feynman diagrams contributing to the
imaginary part of the polarizability. A photon
is absorbed from a bath (denoted by the external crosses),
while a second photon of equal frequency
(nonresonant with respect to an atomic transition)
is emitted (Cutkosky rules).}
\end{minipage}
\end{center}
\end{figure}

By contrast, when the ac Stark shift of an 
atomic level is formulated perturbatively
and the second-order shift of the atomic level in the 
external laser field is evaluated using 
a second-quantized formalism
(see Sec.~III of Ref.~\cite{HaJeKe2006}), a resonance condition
has to be fulfilled in order for an imaginary part of the 
energy shift to be generated. Namely, the final state of
atom$+$field in the decay process
has to have exactly the same energy as the reference 
state of atom$+$field. This is possible 
only at exact resonance, when the emitted photon 
has just the right frequency to compensate the ``quantum
jump'' of the bound electron from an excited state to 
an energetically lower state~\cite{Sa1994Mod,Sa1967Adv,HaJeKe2006}. 
The ac Stark shift is proportional
to the atomic polarizability.
Its tree-level imaginary part~\cite{JeLaDKPa2015,JePa2015epjd1} 
corresponds to spontaneous 
emission of the atom at an exact resonance frequency,
still, not necessarily along the same direction as the 
incident laser photon.
When quantum electrodynamics is involved, it is 
seen that due to quantum fluctuations of the 
electromagnetic field, spontaneous 
emission is possible off resonance.
In Refs.~\cite{JeLaDKPa2015,JePa2015epjd1}, the imaginary part of the 
polarizability was found to be dominated by a self-energy correction
to the ac Stark shift.
Physically, the imaginary part of the 
polarizability corresponds to a ``decay rate'' of the 
reference state $| \phi, n_L \rangle$ used in the 
calculation of the ac Stark shift, to a state
$| \phi, n_L - 1, 1_{\vec k\lambda} \rangle$,
where $| \phi \rangle$ is the atomic reference state,
the occupation number of the laser mode is $n_L$, and there is
either zero or one photon in the mode $\vec k\lambda$.
While the laser frequency is equal to the 
frequency of the emitted radiation ($\omega_L = \omega_{\vec k}$),
the emission proceeds into a different direction as compared 
to the laser wave vector ($\vec k \neq \vec k_L$).
Off resonance, the quantum electrodynamic one-loop effect calculated in 
Refs.~\cite{JeLaDKPa2015,JePa2015epjd1} thus dominates the imaginary part 
of the polarizability, not the tree-level term. This is quite surprising;
the relevant Feynman diagrams are shown in Fig.~\ref{fig1}.
The peculiar behavior of the imaginary 
part of the polarizability suggests a detailed numerical study
of the noncontact friction integral~\cite{ToWi1997,VoPe2002},
and comparison, of the direct and backaction terms.

This paper is organized as follows.  In Sec.~\ref{sec2}, we attempt to shed
some light on the derivation of the effect. 
Full SI~mksA units are kept throughout the derivation.
The numerical calculations of noncontact friction for the hydrogen and helium
interactions with $\alpha$-quartz, gold, and CaF${}_2$ are described in
Sec.~\ref{sec3}, where we shall 
use atomic units for for frequency data and friction coefficients
in Tables~\ref{table1}---\ref{table5}. 
We employ a convenient fit to the vibrational and interband
excitations of the $\alpha$-quartz and CaF${}_2$ lattices.  Finally,
conclusions are drawn in Sec.~\ref{sec4}.

%
%
\section{Derivation}
\label{sec2}

Our derivation is in part inspired by Ref.~\cite{VoPe2002};
we supplement the discussion with some explanatory remarks
and simplified formulas where appropriate.
The electric field at the position of the atomic dipole 
(i.e., at the position of the atom) is written as
\begin{equation}
\vec E(t) = \vec E_0 \, \ee^{-\ii \, \omega \, t} +
\vec E_1 \, \ee^{-\ii \, (\omega + \omega_0) \, t} \,,
\end{equation}
where $\omega$ is the angular frequency component of the (thermal)
fluctuation, and $\omega_0$ describes a small displacement
of the atom's position itself.
The contribution proportional to $\vec E_1$ 
is included as a result of a backaction term,
which takes the variation of the spontaneous and 
induced fields over the spatial amplitude of the oscillatory motion of the 
atom into account [see Eq.~\eqref{backactionE1}].
Hence, the angular frequency of the motion ($\omega_0$)
is added to the thermal frequency,
and the term is proportional to $\exp[-\ii  (\omega + \omega_0) \, t]$.
The displacement of the atom is of angular frequency $\omega_0$,
\begin{equation}
\vec u(t) = \vec u_0 \, \ee^{-\ii \omega_0 t} \,,
\qquad
\vec r(t) = \vec r_0 + \vec u(t) \,.
\end{equation}
The dipole density of the isolated atom is supposed to
perform oscillations of the form
\begin{align}
\label{three}
\vec \pp(\vec r, t) =& \;
\vec \pp_0 \, \delta^{(3)}(\vec r - \vec r_0) \, \ee^{-\ii \omega t}
+ \vec \ppp_1(\vec r, \omega) \, \ee^{-\ii (\omega + \omega_0) \, t} \,,
\nonumber\\
\vec p_1(\vec r, \omega) =& \;
\vec d_1 \, \delta^{(3)}(\vec r - \vec r_0) -
\vec d_0 \, \vec u_0 \cdot 
\vec\nabla \delta^{(3)}(\vec r - \vec r_0) \,.
\end{align}
Here, the second term is generated by the displacement
of the atom,  i.e., by the expansion of the 
Dirac $\delta$ function $\delta^{(3)}(\vec r - \vec r_0 - \vec u(t))$
to first order in $\vec u(t)$.
While the atomic dipole moment is 
a sum of a fluctuating term $\vec \pp^f$ and an 
induced term (by the corresponding frequency component of 
the electric field at the position of the atom),
\begin{equation}
\label{p0}
\pp_{0i} = \pp^f_i + \alpha(\omega) \, E_{0i} \,,
\end{equation}
the frequency component for $\omega + \omega_0$ 
only contains an induced term,
$\vec \pp_1 = \alpha(\omega + \omega_0) \, \vec E_1$.

\begin{figure}[th]
\begin{center}
\begin{minipage}{0.99\linewidth}
\includegraphics[width=0.6\linewidth]{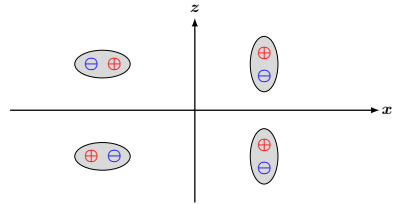}
\caption{\label{fig2} (Color online.)
Mirroring a dipole in the $xy$ plane. A dipole aligned 
along the $x$ axis gives rise to an antiparallel mirror dipole,
whereas a dipole aligned along the $z$ axis gives
rise to a parallel mirror dipole.
Recall that mirror charges have the opposite sign as compared 
to the original ones.}
\end{minipage}
\end{center}
\end{figure}

Let $\GG_{ij}(\vec r, \vec r_0, \omega)$
denote the frequency component of the 
Green tensor which determines the electric field
generated at position $\vec r$ by a point dipole at $\vec r_0$.
In the nonretardation approximation [Eq.~(1) of Ref.~\cite{ToWi1997}],
it reads
\begin{align}
g(\vec r, \vec r', \omega) =& \;
\frac{1}{4 \pi \epsilon_0} \,
\left( \frac{1}{|\vec r - \vec r'|} \right.
\nonumber\\[0.133ex]
& \; \left. - \frac{\epsilon(\omega)-1}{\epsilon(\omega)+1} \,
\frac{1}{|\vec r - \vec r' + 
2 \hat n_\perp (\vec r' \cdot \hat n_\perp)|} \right) \,,
\nonumber\\[0.133ex]
\GG_{ij}(\vec r, \vec r', \omega) =& \;
-\nabla_i \nabla'_j \, g(\vec r, \vec r', \omega) \,.
\end{align}
Here, $\hat n = \hat e_z$ is the surface normal (the 
surface of the dielectric is the $xy$ plane).
The result
\begin{equation}
\label{second_term}
\GG_{zz}(\vec 0, \vec r_\calZ, \omega) =
\frac{2}{\calZ^3} + 
\frac{\epsilon(\omega) - 1}{\epsilon(\omega)+1} 
\frac{2}{\calZ^3}\,,
\quad
\vec r_\calZ = \hat e_z \, \calZ \,,
\end{equation}
reflects the fact that a dipole oriented in parallel to the $z$ axis
generates a mirror dipole which also is oriented in parallel to the $z$
axis (not antiparallel, see the dipoles in Fig.~\ref{fig2}).
Because of this, the second term on the right-hand side
of Eq.~\eqref{second_term} has the same sign as the first term.

Self-consistency dictates that the 
field $\vec E_0 \equiv \vec E_0(\vec r_0)$ at the position of the atom is 
equal to the sum of the field generated by 
the dipole moment $\pp_{0i}$, and the fluctuating 
component $E^s_i(\vec r_0, \omega)$ of the electric field,
\begin{align}
E_{0i}
=& \; \GG_{ii}(\vec r_0, \vec r_0, \omega) \, \pp_{0i}
+ E^s_i(\vec r_0, \omega) 
\nonumber\\[0.133ex]
=& \; 
\GG_{ii}(\vec r_0, \vec r_0, \omega) \,
\alpha(\omega) E_{0i} + 
\GG_{ii}(\vec r_0, \vec r_0, \omega) \, \pp^f_i 
\nonumber\\[0.133ex]
& \; + E^s_i(\vec r_0, \omega) \,,
\end{align}
where no summation over $i$ is carried out
(one has $\GG_{ij} = G_{ii} \, \delta_{ij}$ at
equal spatial coordinates). So,
\begin{subequations}
\label{back}
\begin{align}
\label{backa}
E_{0i} =& \; \frac{ \GG_{ii}(\vec r_0, \vec r_0, \omega) \, \pp^f_i +
E^s_i(\vec r_0, \omega) }%
{ 1 - \GG_{ii}(\vec r_0, \vec r_0, \omega) \, \alpha(\omega) } \,,
\\[0.133ex]
\label{backb}
\pp_{0i} =& \; 
\frac{ \pp^f_i + \alpha(\omega) \, E^s_i(\vec r_0, \omega) }%
{1 - \alpha(\omega) \, \GG_{ii}(\vec r_0, \vec r_0, \omega)} \,,
\end{align}
\end{subequations}
where in Eq.~\eqref{backb} we have taken into account Eq.~\eqref{p0}.
The electric field $\vec E_{0}$ and the dipole moment $\vec \pp_0$
are given in terms of fluctuating terms;
the denominators in Eq.~\eqref{back} take the backaction into account.
For $\vec E_1$, one observes that the gradient term in the expression
of $\vec p_1(\vec r, \omega)$ [Eq.~\eqref{three}], 
in the non-fluctuating contribution $\int \dd^3 r' \,
\GG_{ij}(\vec r, \vec r', \omega+\omega_0) \, p_{1j}(\vec r, \omega)$,
needs to be treated by partial integration. Adding the term due to
the fluctuations of the atom's position, and due to the spontaneous
fluctations of the electromagnetic field, one obtains
\begin{align}
\label{backactionE1}
E_{1i} =& \; \GG_{ii}(\vec r_0, \vec r_0, \omega + \omega_0) \,
\alpha(\omega + \omega_0) \, E_{1i}
\nonumber\\[0.133ex]
& \;
+ \vec u_0 \cdot \vec\nabla_{\vec r} \left( E^s_i(\vec r, \omega)
+ \GG_{ij}(\vec r_0, \vec r, \omega + \omega_0) \, \pp_{0j}
\right.
\nonumber\\[0.133ex]
& \; \left. \left.
+ \GG_{ij}(\vec r, \vec r_0, \omega) \, \pp_{0j} \right)
\right|_{\vec r = \vec r_0} \,.
\end{align}
This equation can be trivially solved for $\vec E_1$. 
The thermal fluctuations are described by the following equations~\cite{ToWi1997},
\begin{subequations}
\begin{align}
\left< \pp^f_i \, \pp^f_j \right>_\omega =& \; 
\frac{2 \, \Theta(\omega, T)}{\omega} \,
\delta_{ij} \, {\rm Im} \, \alpha(\omega) \,,
\\[0.133ex]
\left< E_i(\vec r) \, E_j(\vec r') \right>_\omega =& \;
\frac{2 \, \Theta(\omega, T)}{\omega} \,
{\rm Im} [G_{ij}(\vec r, \vec r', \omega)]  \,.
\end{align}
\end{subequations}
where $\Theta(\omega, T) = 
\hbar \, \omega \, 
\left( \tfrac12 + n(\omega) \right)
= \tfrac12 \hbar\, \omega \,
\coth\left( \tfrac12 \, \beta \, \hbar \, \omega \right)$ 
is the Kallen--Welton thermal factor, with
$n(\omega) = [\exp(\beta\, \hbar\, \omega) - 1]^{-1}$,
and $\beta = 1/(k_B \, T)$ where $k_B$
is the Boltzmann constant.
With the help of $\rho = -\vec \nabla \cdot \vec p$
and $\vec j = \partial_t \vec p$, one formulates a 
time-dependent force,
\begin{align}
\vec F(t) =& \;
\int \dd^3 r \, 
\left< \rho(\vec r, t) \, \vec E^*(\vec r, t) 
+ \vec j(\vec r, t) \times \vec B^*(\vec r, t)
\right>
\nonumber\\[0.133ex]
=& \; \vec F_s(t) + 
\vec u_0 \cdot \frac{\partial}{\partial r} \vec F_s(t) +
\vec F_f(\omega, \omega_0) \, \ee^{-\ii \omega_0 \, t} \,.
\end{align}
Here, $F_s(t)$ is the static van-der-Waals force, 
$\vec u_0 \cdot \frac{\partial}{\partial r} \vec F_s(t)$
describes the variation of the van-der-Waals force
with the oscillating position of the atom, 
and $\vec F_f(\omega, \omega_0)$ is a Fourier
component of the friction force.
An integration over the thermal fluctuations of all 
Fourier components of the friction force gives the
total friction force,
\begin{align}
\vec F_f =& \; 
\frac12 \, \left. 
\int_0^\infty \frac{\dd \omega}{2 \pi} \;
\omega_0 \, \frac{\partial}{\partial \omega_0} 
\left< \vec F(\omega, \omega_0) \right>
\right|_{\omega_0 = 0} 
\nonumber\\[0.133ex]
=& \; \ii \, \omega_0 \, 
\left[ \eta_x \, (u_{0x} \hat e_x + u_{0y} \hat e_y) +
\eta_z \, u_z \hat e_z \right] 
\nonumber\\[0.133ex]
=& \; -\eta_x \, (v_x \hat e_x + v_y \hat e_y) - 
\eta_z v_z \hat e_z \,.
\end{align}
Here, $\eta_x$ and $\eta_z$ are the friction coefficient 
for motion along the $x$ and $z$ directions, respectively.
The additional assumption of a small mechanical motion
with velocity $\vec v = \partial_t \left. \vec u_0 \ee^{-\ii \omega_0 t} 
\right|_{t = 0} = -\ii \omega_0 \, \vec u_0$ is made.

The result for $\eta_x$ is obtained as,
\begin{widetext}
\begin{align}
\label{commensurate}
\eta_x =& \; 
\frac{\beta \hbar^2 }{2 \pi} \, \int_0^\infty 
\frac{\dd \omega}{\sinh^2( \tfrac12 \, \beta \, \hbar \, \omega )} \,
\left[ \sum_{\ell = x,y,z}
\frac{\partial^2}{\partial x \, \partial x'}
\im \GG_{\ell \ell}(\vec r, \vec r', \omega) \,
\im \left( \frac{ \alpha(\omega) }{1 - \alpha(\omega) \,
\GG_{\ell \ell}(\vec r_\calZ, \vec r_\calZ, \omega) } \right)
\right.
\nonumber\\[0.133ex]
& \; \left. \left. - 2 |\alpha(\omega)|^2 \, \re \left(
\frac{1}{ (1- \alpha^*(\omega) \, D^*_{zz}(\vec r_\calZ, \vec r_\calZ, \omega))
(1- \alpha(\omega) \, \GG_{zz}(\vec r_\calZ, \vec r_\calZ, \omega))} \right) \,
\left( \frac{\partial}{\partial x} \GG_{xz}(\vec r, \vec r_\calZ, \omega) \right)^2
\right] \right|_{\vec r, \vec r' = \vec r_\calZ }
\nonumber\\[0.133ex]
\approx & \;
\frac{\beta \hbar^2}{2 \pi} \, \int_0^\infty 
\frac{\dd \omega}{\sinh^2( \tfrac12 \, \beta \, \hbar \, \omega )} \,
\left[ \sum_{\ell = x,y,z} \frac{\partial^2}{\partial x \, \partial x'}
\im[\GG_{\ell \ell}(\vec r, \vec r', \omega)]\, \im[\alpha(\omega)]
+ \alpha(\omega)^2 
\right.
\nonumber\\[0.133ex]
& \; \left. \left. \times \left\{
\sum_{\ell = x,y,z} \left\{ \frac{\partial^2}{\partial x \, \partial x'}
\im [ \GG_{\ell \ell}(\vec r, \vec r', \omega) ] \,
\im [ \GG_{\ell \ell}(\vec r_\calZ, \vec r_\calZ, \omega) ]
\right\}
- 2  \left( \frac{\partial}{\partial x}
\im \left[ \GG_{x z}(\vec r, \vec r_\calZ, \omega) \right] \right)^2
\right\} 
\right]
\right|_{\vec r, \vec r' = \vec r_\calZ } \,.
\end{align}
This result can be written as $\eta_x = \eta_x^{(1)} + \eta_x^{(2)}$,
where $\eta_x^{(2)}$ is generated by the term in curly brackets in the 
integrand.  With the help of
$\sum_\ell \frac{\partial^2}{\partial x \, \partial x'}
\im \, \GG_{\ell \ell}(\vec r, \vec r') =
\im \left( \frac{\epsilon(\omega) - 1}{\epsilon(\omega) + 1} \right) \,
\frac{3}{16 \pi \epsilon_0 \, \calZ^5}$,
one verifies that the leading-order, linear term in the polarizability
(see Ref.~\cite{ToWi1997}),
from Eq.~\eqref{commensurate}, is given as
\begin{align}
\label{eta1X}
\eta^{(1)}_x = & \;
\frac{\beta \hbar^2 }{2 \pi} \, \int_0^\infty 
\frac{\dd \omega}{\sinh^2( \tfrac12 \, \beta \, \hbar \, \omega )} 
\sum_{\ell = x,y,z}
\frac{\partial^2}{\partial x \, \partial x'}
\im \GG_{\ell \ell}(\vec r, \vec r') \,
\im \left( \alpha(\omega) \right)
= \frac{3 \beta \hbar^2}{32 \pi^2 \epsilon_0 \calZ^5} \, 
\int_0^\infty
\frac{\dd \omega \, 
\im [ \alpha(\omega) ]}{\sinh^2( \tfrac12 \, \beta \, \hbar \, \omega )} 
\im \left( \frac{\epsilon(\omega) - 1}{\epsilon(\omega) + 1} \right).
\end{align}
In Eq.~\eqref{commensurate},
the term of second order in the polarizability is given as follows,
\begin{align}
\label{eta2X}
\eta^{(2)}_x = & \;
\left. \frac{\beta \hbar^2 }{8 \pi} \, \int\limits_0^\infty 
\frac{\dd \omega \,\alpha(\omega)^2}{\sinh^2( \tfrac12 \, \beta \, \hbar \, \omega )} \,
\left[ \left\{ \frac{\partial^2}{\partial z^2}
\im \, \GG_{zz}(\vec r, \vec r_\calZ, \omega) \right\} \,
\im \, \GG_{zz}(\vec r_\calZ, \vec r_\calZ, \omega) 
- 2 \, 
\left( \frac{\partial}{\partial z} \im \, 
\GG_{zz}(\vec r, \vec r_\calZ, \omega) \right)^2
\right] \right|_{\vec r, \vec r' = \vec r_\calZ }
\nonumber\\[0.133ex]
=& \; \frac{9 \beta \hbar^2}{4096\, \pi^3 \, \epsilon_0^2 \, \calZ^8} \,
\int\limits_0^\infty \dd \omega \,
\frac{\alpha(\omega)^2}{\sinh^2( \tfrac12 \, \beta \, \hbar \, \omega)}
\left[ \im \left( \frac{\epsilon(\omega) - 1}{\epsilon(\omega) + 1} \right)
\right]^2 \,.
\end{align}
\end{widetext}
For friction in the $z$ direction, one derives 
$\eta_z = \eta_z^{(1)} + \eta_z^{(2)}$,
with $\eta_z^{(1)} = 2 \, \eta^{(2)}_z$ and 
$\eta_z^{(2)} = 7 \, \eta^{(2)}_x$, confirming Ref.~\cite{VoPe2002}.
The term $\eta^{(2)}$ is generated by the 
``backaction denominators'' from Eqs.~\eqref{backa} and~\eqref{backb}.
For the numerical evaluation of the 
term $\eta^{(1)}$, the following result 
\begin{subequations}
\label{mainres}
\begin{align}
\label{mainres1}
{\rm Im}[ \alpha(\omega) ] =& \;
{\rm Im}[ \alpha_R(\omega) ] +
\frac{\omega^3}{6 \pi \epsilon_0 c^3} \, [ \alpha(\omega) ]^2 \,,
\\[0.133ex]
{\rm Im}\left[ \alpha_R(\omega) \right] =& \;
{\rm Im}\left[ \alpha_r(\omega) \right] -
{\rm Im}\left[ \alpha_r(-\omega) \right] \,,
\\[0.133ex]
\label{mainres2}
{\rm Im}\left[ \alpha_r(\omega) \right] =& \;
\frac{\pi}{2} \, \sum_m \frac{f_{m0}}{E_m - E} \,
\delta(E_m - E + \hbar \, \omega) \,,
\end{align}
\end{subequations}
has recently been derived in Ref.~\cite{JeLaDKPa2015}.
Here, $f_{m0}$ are the oscillator strengths~\cite{BeSa1957,YaBaDaDr1996}
for the dipole transitions from the ground state of the atom
with energy $E$ to the excited states $| m \rangle$ with 
energy $E_m$. The ``one-loop'' term in the 
result for ${\rm Im}[ \alpha(\omega) ]$,
proportional to $\alpha(\omega)^2$, 
implies that the numerical evaluation of 
both $\eta^{(1)}$ and $\eta^{(2)}$ is related;
because typical thermal wave vectors 
(inversely related to the thermal wavelengths) are much 
smaller than typical atomic transition frequencies, 
$\eta^{(2)}$ is the dominant term.
The resonant, tree-level contribution to the 
atomic polarizability is denoted as ${\rm Im}\left[ \alpha_r(\omega) \right]$.

The expression for ${\rm Im}\left[ \alpha_r(\omega) \right]$ 
takes into account only resonant
processes, with Dirac-$\delta$ peaks near the resonant transitions.  However,
this concept ignores the possibility of off-resonant driving of an atomic
transition, where the atom would absorb an off-resonant photon and emit a
photon of the same frequency as the absorbed, off-resonant one, but in a
different spatial direction. Indeed, it has been argued in
Ref.~\cite{LaDKJe2012prl} that the off-resonant driving of an atomic transition
mediates the dominant mechanism in the determination of the quantum friction
force.  The same argument applies to the atom-surface quantum friction force
mediated by the dragging of the image dipole inside the medium, which is the
subject of the current investigation.  We have recently considered (see
Ref.~\cite{JePa2015epjd1}) the Feynman diagrams in Fig.~\ref{fig1},
where the ``grounded'' external photon lines (those ``anchored'' by the 
external crosses) represent the absorption of an off-resonant
photon from the quantized radiation field 
(e.g., a laser field or a bath of thermal photons), 
the vertical internal lines denote the ``cutting'' of the diagram
at the point where the photon is emitted, 
and the photon loop denotes the self-interaction of the 
atomic electron (the imaginary of the corresponding 
energy shift is directly proportional to the imaginary 
part of the polarizability~\cite{BaSu1978}). 
The overall result is obtained by adding the (in this case dominant) one-loop 
``correction'' to the resonant imaginary part of the 
polarizability.

\begin{table}[t]
\caption{\label{table1} Coefficients for the first few resonances for
$\alpha$-quartz according to the fitting formula~\eqref{fit_formula} (ordinary
and extraordinary optical axes). 
The $\omega_k$ and $\gamma_k$ are measured in atomic units,
i.e., in units of the $E_h/\hbar$, where $E_h$ is the Hartree energy.  The
fitting parameters have been obtained from data tabulated in Ref.~\cite{Pa1985}
(see also Ref.~\cite{LaDKJe2010pra}).}
\begin{center}
\begin{tabular}{c @{\hspace*{0.5cm}} D{,}{.}{11} D{,}{.}{11} D{,}{.}{11}}
\hline
\hline
\multicolumn{4}{c}{\stdrule Vibrational Excitations (Ordinary Axis)} \\
\stdrule
$k$
& \multicolumn{1}{c}{$\alpha_k$}
& \multicolumn{1}{c}{$\omega_k$}
& \multicolumn{1}{c}{$\gamma_k$} \\
\hline
\stdrule
 1  & 1,04 \times 10^{-2} & 1,83 \times 10^{-3} & 1,29 \times 10^{-5} \\
\stdrule
 2  & 8,53 \times 10^{-2} & 2,22 \times 10^{-3} & 1,83 \times 10^{-5} \\
\stdrule
 3  & 0,16 \times 10^{-2} & 3,18 \times 10^{-3} & 3,16 \times 10^{-5} \\
\stdrule
 4  & 1,06 \times 10^{-2} & 3,67 \times 10^{-3} & 3,20 \times 10^{-5} \\
\stdrule
 5  & 5,52 \times 10^{-2} & 5,23 \times 10^{-3} & 3,61 \times 10^{-5} \\
\stdrule
 6  & 4,55 \times 10^{-2} & 5,34 \times 10^{-3} & 3,89 \times 10^{-5} \\
\hline
\hline
\multicolumn{4}{c}{\stdrule Interband Excitations (Ordinary Axis)} \\
\stdrule
$k$
& \multicolumn{1}{c}{$\alpha_k$}
& \multicolumn{1}{c}{$\omega_k$}
& \multicolumn{1}{c}{$\gamma_k$} \\
\hline
\stdrule
 7  & 1,05 \times 10^{-2} & 3,89 \times 10^{-1} & 1,12 \times 10^{-2} \\
\stdrule
 8  & 4,71 \times 10^{-2} & 4,45 \times 10^{-1} & 5,28 \times 10^{-2} \\
\stdrule
 9  & 4,98 \times 10^{-2} & 5,37 \times 10^{-1} & 7,32 \times 10^{-2} \\
\stdrule
10  & 1,06 \times 10^{-1} & 6,58 \times 10^{-1} & 1,30 \times 10^{-1} \\
\stdrule
11  & 1,12 \times 10^{-1} & 8,26 \times 10^{-1} & 2,40 \times 10^{-1} \\
\hline
\hline
\multicolumn{4}{c}{\stdrule Vibrational Excitations (Extraordinary Axis)} \\
\stdrule
$k$
& \multicolumn{1}{c}{$\alpha_k$}
& \multicolumn{1}{c}{$\omega_k$}
& \multicolumn{1}{c}{$\gamma_k$} \\
\hline
\hline
\stdrule
 1 & 3,63 \times 10^{-2} & 1,74 \times 10^{-3} & 2,32 \times 10^{-5} \\
\stdrule
 2 & 8,45 \times 10^{-4} & 2,31 \times 10^{-3} & 1,52 \times 10^{-5} \\
\stdrule
 3 & 7,54 \times 10^{-2} & 2,42 \times 10^{-3} & 3,00 \times 10^{-5} \\
\stdrule
 4 & 1,08 \times 10^{-2} & 3,58 \times 10^{-3} & 3,49 \times 10^{-5} \\
\stdrule
 5 & 1,03 \times 10^{-1} & 5,31 \times 10^{-3} & 4,46 \times 10^{-5} \\
\hline
\multicolumn{4}{c}{\stdrule Interband Excitations (Extraordinary Axis)} \\
\stdrule
$k$
& \multicolumn{1}{c}{$\alpha_k$}
& \multicolumn{1}{c}{$\omega_k$}
& \multicolumn{1}{c}{$\gamma_k$} \\
\hline
\stdrule
 6 & 1,05 \times 10^{-2} & 3,89 \times 10^{-1} & 1,12 \times 10^{-2} \\
\stdrule
 7 & 4,71 \times 10^{-2} & 4,45 \times 10^{-1} & 5,28 \times 10^{-2} \\
\stdrule
 8 & 4,98 \times 10^{-2} & 5,37 \times 10^{-1} & 7,32 \times 10^{-2} \\
\stdrule
 9 & 1,06 \times 10^{-1} & 6,58 \times 10^{-1} & 1,30 \times 10^{-2} \\
\stdrule
10 & 1,12 \times 10^{-1} & 8,26 \times 10^{-1} & 2,40 \times 10^{-2} \\
\hline
\hline
\end{tabular}
\end{center}
\end{table}

\begin{table}[t]
\caption{\label{table2} Same as Table~\ref{table1} but the 
data are for CaF${}_2$. The fitting parameters 
are obtained using numerical data compiled in 
Refs.~\cite{OrEtAl1983,OrEtAl1985,Pa1985,KaEtAl1962,DeEtAl1970,PSEtAl2009}
for the optical response function of CaF${}_2$.}
\begin{center}
\begin{tabular}{c @{\hspace*{0.5cm}} 
D{,}{.}{11} D{,}{.}{11} D{,}{.}{11}}
\hline
\hline
\multicolumn{4}{c}{\stdrule Vibrational Excitations (CaF${}_2$)} \\
\stdrule
$k$
& \multicolumn{1}{c}{$\alpha_k$}
& \multicolumn{1}{c}{$\omega_k$}
& \multicolumn{1}{c}{$\gamma_k$} \\
\hline
\stdrule
 1 & 4,25  \times 10^{-1} & 1,74 \times 10^{-3} & 1,49 \times 10^{-4} \\
\hline
\multicolumn{4}{c}{\stdrule Interband Excitations (CaF${}_2$)} \\
\stdrule
$k$
& \multicolumn{1}{c}{$\alpha_k$}
& \multicolumn{1}{c}{$\omega_k$}
& \multicolumn{1}{c}{$\gamma_k$} \\
\hline
\stdrule
 2 & 9,85 \times 10^{-3} & 4,12 \times 10^{-1} & 1,98 \times 10^{-2} \\
\stdrule
 3 & 1,62 \times 10^{-1} & 5,74 \times 10^{-1} & 1,72 \times 10^{-1} \\
\stdrule
 4 & 1,57 \times 10^{-1} & 1,13 \times 10^{0}  & 5,58 \times 10^{-1} \\
\hline
\hline
\end{tabular}
\end{center}
\end{table}

%
%
\begin{table*}[t]
\begin{center}
\begin{minipage}{0.95\linewidth}
\begin{center}
\caption{\label{table3} Normalized friction coefficients
$\eta^{(1)}_{0x}$ and $\eta^{(2)}_{0x}$, given in atomic units (denoted as a.u.),
for a distance of $\calZ = a_0$ from the 
$\alpha$-quartz surface, obtained using the 
expression~\eqref{ImAlphaAU} for the imaginary part of the atomic polarizability
and using Eqs.~\eqref{eta1} and~\eqref{eta2} for the friction coefficients.
The friction coefficient, in SI~mksA units, is obtained from Eqs.~\eqref{eta_au}
and~\eqref{eta_given}.}
\begin{tabular}{c@{\hspace{1.5cm}}l@{\hspace{0.8cm}}l@{\hspace{0.8cm}}l%
@{\hspace{0.8cm}}l@{\hspace{0.8cm}}l@{\hspace{0.8cm}}l}
\hline
\hline
    \multicolumn{7}{c}{\rule[-2mm]{0mm}{6mm} 
    Friction Coefficients for SiO${}_2$ [Ordinary Axis]}\\
\rule[-2mm]{0mm}{6mm}
    & \multicolumn{2}{c}{Atomic Hydrogen ($1S$)}
    & \multicolumn{2}{c}{Helium ($1S$)} 
    & \multicolumn{2}{c}{Helium ($2 {}^3S_1$)} \\
\rule[-2mm]{0mm}{6mm}
$T$ [K] 
    & \multicolumn{1}{l}{$\eta^{(1)}_{x0}$}
    & \multicolumn{1}{l}{$\eta^{(2)}_{x0}$}
    & \multicolumn{1}{l}{$\eta^{(1)}_{x0}$}
    & \multicolumn{1}{l}{$\eta^{(2)}_{x0}$}
    & \multicolumn{1}{l}{$\eta^{(1)}_{x0}$}
    & \multicolumn{1}{l}{$\eta^{(2)}_{x0}$} \\
\hline
\rule[-1mm]{0mm}{4mm}
273 &  $2.05 \times 10^{-15}$ 
    &  $1.76 \times 10^{-1}$ 
    &  $1.94  \times 10^{-16}$ 
    &  $1.67 \times 10^{-2}$
    &  $1.03 \times 10^{-11}$ 
    &  $8.75 \times 10^{2}$ \\
\rule[-1mm]{0mm}{4mm}
298 &  $2.78 \times 10^{-15}$ 
    &  $2.14 \times 10^{-1}$ 
    &  $2.63 \times 10^{-16}$ 
    &  $2.02 \times 10^{-2}$
    &  $1.40 \times 10^{-11}$ 
    &  $1.06 \times 10^{3}$ \\
\rule[-1mm]{0mm}{4mm}
300 &  $2.85 \times 10^{-15}$ 
    &  $2.17 \times 10^{-1}$ 
    &  $2.69 \times 10^{-16}$ 
    &  $2.05 \times 10^{-2}$
    &  $1.43 \times 10^{-11}$ 
    &  $1.08 \times 10^{3}$ \\
\hline
\hline
   \multicolumn{7}{c}{\rule[-2mm]{0mm}{6mm} Friction Coefficients for SiO${}_2$ 
      [Extraordinary Axis]}\\
\rule[-2mm]{0mm}{6mm}
    & \multicolumn{2}{c}{Atomic Hydrogen ($1S$)}
    & \multicolumn{2}{c}{Helium ($1S$)}
    & \multicolumn{2}{c}{Helium ($2 {}^3S_1$)} \\
\rule[-2mm]{0mm}{6mm}
$T$ [K]
    & \multicolumn{1}{l}{$\eta^{(1)}_{x0}$}
    & \multicolumn{1}{l}{$\eta^{(2)}_{x0}$}
    & \multicolumn{1}{l}{$\eta^{(1)}_{x0}$}
    & \multicolumn{1}{l}{$\eta^{(2)}_{x0}$}
    & \multicolumn{1}{l}{$\eta^{(1)}_{x0}$}
    & \multicolumn{1}{l}{$\eta^{(2)}_{x0}$} \\
\hline
\rule[-1mm]{0mm}{4mm}
273 &  $2.00 \times 10^{-15}$
    &  $9.19 \times 10^{-2}$
    &  $1.89 \times 10^{-16}$
    &  $1.67 \times 10^{-2}$
    &  $1.01 \times 10^{-11}$
    &  $4.57 \times 10^{2}$ \\
\rule[-1mm]{0mm}{4mm}
298 &  $2.70 \times 10^{-15}$
    &  $1.14 \times 10^{-1}$
    &  $2.55 \times 10^{-16}$
    &  $2.02 \times 10^{-2}$
    &  $1.36 \times 10^{-11}$
    &  $5.69 \times 10^{2}$ \\
\rule[-1mm]{0mm}{4mm}
300 &  $2.76 \times 10^{-15}$
    &  $1.16 \times 10^{-1}$
    &  $2.61 \times 10^{-16}$
    &  $2.05 \times 10^{-2}$
    &  $1.39 \times 10^{-11}$
    &  $5.78 \times 10^{2}$ \\
\hline
\hline
\end{tabular}
\end{center}
\end{minipage}
\end{center}
\end{table*}

\begin{table*}[t]
\begin{center}
\begin{minipage}{0.95\linewidth}
\begin{center}
\caption{\label{table4} Same as Table~\ref{table3},
but for the hydrogen and helium interactions with gold (Au).}
\begin{tabular}{c@{\hspace{1.5cm}}l@{\hspace{0.8cm}}l@{\hspace{0.8cm}}l%
@{\hspace{0.8cm}}l@{\hspace{0.8cm}}l@{\hspace{0.8cm}}l}
\hline
\hline
    \multicolumn{7}{c}{\rule[-2mm]{0mm}{6mm} Friction Coefficients for Gold (Au)}\\
\rule[-2mm]{0mm}{6mm}
    & \multicolumn{2}{c}{Atomic Hydrogen ($1S$)}
    & \multicolumn{2}{c}{Helium ($1S$)} 
    & \multicolumn{2}{c}{Helium ($2 {}^3S_1$)} \\
\rule[-2mm]{0mm}{6mm}
$T$ [K] 
    & \multicolumn{1}{l}{$\eta^{(1)}_{x0}$}
    & \multicolumn{1}{l}{$\eta^{(2)}_{x0}$}
    & \multicolumn{1}{l}{$\eta^{(1)}_{x0}$}
    & \multicolumn{1}{l}{$\eta^{(2)}_{x0}$}
    & \multicolumn{1}{l}{$\eta^{(1)}_{x0}$}
    & \multicolumn{1}{l}{$\eta^{(2)}_{x0}$} \\
\hline
\rule[-1mm]{0mm}{4mm}
273 &  $8.67 \times 10^{-19}$ 
    &  $1.05 \times 10^{-9}$ 
    &  $8.19 \times 10^{-20}$ 
    &  $9.91 \times 10^{-11}$ 
    &  $4.38 \times 10^{-15}$ 
    &  $5.20 \times 10^{-6}$ \\
\rule[-1mm]{0mm}{4mm}
298 &  $1.26 \times 10^{-15}$ 
    &  $1.27 \times 10^{-9}$ 
    &  $1.19 \times 10^{-19}$ 
    &  $1.20 \times 10^{-10}$ 
    &  $6.41 \times 10^{-15}$ 
    &  $6.32 \times 10^{-6}$ \\
\rule[-1mm]{0mm}{4mm}
300 &  $1.30 \times 10^{-15}$ 
    &  $1.29 \times 10^{-9}$ 
    &  $1.23 \times 10^{-19}$ 
    &  $1.22 \times 10^{-10}$ 
    &  $6.60 \times 10^{-15}$ 
    &  $6.41 \times 10^{-6}$ \\
\hline
\hline
\end{tabular}
\end{center}
\end{minipage}
\end{center}
\end{table*}

\begin{table*}[t]
\begin{center}
\begin{minipage}{0.95\linewidth}
\begin{center}
\caption{\label{table5} Same as Table~\ref{table3},
but for the hydrogen and helium interactions with CaF${}_2$.}
\begin{tabular}{c@{\hspace{1.5cm}}l@{\hspace{0.8cm}}l@{\hspace{0.8cm}}l%
@{\hspace{0.8cm}}l@{\hspace{0.8cm}}l@{\hspace{0.8cm}}l}
\hline
\hline
    \multicolumn{7}{c}{\rule[-2mm]{0mm}{6mm} Friction Coefficients for CaF${}_2$}\\
\rule[-2mm]{0mm}{6mm}
    & \multicolumn{2}{c}{Atomic Hydrogen ($1S$)}
    & \multicolumn{2}{c}{Helium ($1S$)} 
    & \multicolumn{2}{c}{Helium ($2 {}^3S_1$)} \\
\rule[-2mm]{0mm}{6mm}
$T$ [K] 
    & \multicolumn{1}{l}{$\eta^{(1)}_{x0}$}
    & \multicolumn{1}{l}{$\eta^{(2)}_{x0}$}
    & \multicolumn{1}{l}{$\eta^{(1)}_{x0}$}
    & \multicolumn{1}{l}{$\eta^{(2)}_{x0}$}
    & \multicolumn{1}{l}{$\eta^{(1)}_{x0}$}
    & \multicolumn{1}{l}{$\eta^{(2)}_{x0}$} \\
\hline
\rule[-1mm]{0mm}{4mm}
273 &  $3.12 \times 10^{-15}$ 
    &  $4.79 \times 10^{-1}$ 
    &  $8.34 \times 10^{-16}$ 
    &  $4.53 \times 10^{-2}$ 
    &  $1.54 \times 10^{-11}$ 
    &  $2.37 \times 10^{3}$ \\
\rule[-1mm]{0mm}{4mm}
298 &  $3.61 \times 10^{-15}$ 
    &  $5.09 \times 10^{-1}$ 
    &  $8.85 \times 10^{-16}$ 
    &  $4.81 \times 10^{-2}$ 
    &  $1.78 \times 10^{-11}$ 
    &  $2.52 \times 10^{3}$ \\
\rule[-1mm]{0mm}{4mm}
300 &  $3.65 \times 10^{-15}$ 
    &  $5.11 \times 10^{-1}$ 
    &  $8.88 \times 10^{-16}$ 
    &  $4.83 \times 10^{-2}$ 
    &  $1.80 \times 10^{-11}$ 
    &  $2.53 \times 10^{3}$ \\
\hline
\hline
\end{tabular}
\end{center}
\end{minipage}
\end{center}
\end{table*}

%
%
\section{Numerical Evaluation}
\label{sec3}

The structure of Eqs.~\eqref{eta1X} and~\eqref{eta2X},
which we recall for convenience,
\begin{subequations}
\begin{align}
\label{eta1}
\eta^{(1)}_x = & \;
\frac{3 \beta \hbar^2}{32 \pi^2 \epsilon_0 \calZ^5} \,
\int_0^\infty
\frac{\dd \omega \,
\im [ \alpha(\omega) ]}{\sinh^2( \tfrac12 \, \beta \, \hbar \, \omega )}
\im \left( \frac{\epsilon(\omega) - 1}{\epsilon(\omega) + 1} \right) \,,
\\
\label{eta2}
\eta^{(2)}_x = & \;
\frac{9 \beta \hbar^2}{4096\, \pi^3 \, \epsilon_0^2 \, \calZ^8} \,
\nonumber\\[0.133ex]
& \; \times \int\limits_0^\infty \dd \omega \,
\frac{\alpha(\omega)^2}{\sinh^2( \tfrac12 \, \beta \, \hbar \, \omega)}
\left[ \im \left( \frac{\epsilon(\omega) - 1}{\epsilon(\omega) + 1} \right)
\right]^2 \,,
\end{align}
\end{subequations}
implies that,
for the evaluation of the quantum friction coefficient in 
the vicinity of a dielectric, we need to have 
reliable data for both the imaginary part of the
polarizability of the atom, ${\rm Im}[\alpha(\omega)]$,
as well as the 
imaginary part of the dielectric response function,
which is given as
${\rm Im}[ (\epsilon(\omega)-1)/(\epsilon(\omega)+1) ]$.
A related problem, namely, the calculation of 
black-body friction for an atom immersed in 
a thermal bath of photons, has recently  been 
considered in Ref.~\cite{LaDKJe2012prl}.
It has been argued that the
inclusion of the width $\Gamma_n$ of the virtual states in the expression for
the polarizability is crucial for obtaining reliable predictions.
The imaginary part of the polarizability is given 
in Eq.~\eqref{mainres}.

In the SI mksA unit system~\cite{MoTaNe2012}, the atomic dipole polarizability
describes the dynamically induced dipole, which is created when the atom is
irradiated with a light field (electric field).
Thus, the physical dimension of the polarizability,
in SI mksA units, is determined by the requirement that
one should obtain a dipole moment upon multiplying the
polarizability $\alpha(\omega)$ by an electric field.
In atomic units (a.u.) with $\hbar = 1$, $c= 1/\alpha$,
and $\epsilon_0 = 1/(4 \pi)$, one has
\begin{equation}
\label{ImAlphaAU}
\left. {\rm Im}[\alpha(\omega)] \right|_{\rm a.u.} =
\left. {\rm Im}[\alpha_R(\omega)] \right|_{\rm a.u.} +
\frac{2 \alpha^3}{3} \,
\left. \left\{ \omega^3 [\alpha(\omega)]^2 \right\} \right|_{\rm a.u.} \,.
\end{equation}

In natural as well as atomic units~\cite{BeSa1957},
physical quantities are identified with the 
corresponding reduced quantities, i.e.,
with the numbers that multiply the fundamental 
units in the respective unit systems.
In order to convert the relation~\eqref{mainres2} into 
atomic units, we recall that the atomic units for 
charge ($e$), length (Bohr radius $a_0$), and energy (Hartree $E_h$)
are as follows,
\begin{subequations}
\begin{align}
|e| =& \; 1.60218 \times 10^{-19} \, {\rm C}\,,
\\[0.133ex]
a_0 =& \; \frac{\hbar}{\alpha \, m_e \, c} = 5.29177 \times 10^{-11} \, {\rm m}\,,
\\[0.133ex]
E_h =& \; m_e \, (\alpha \, c)^2 = 4.35974 \times 10^{-18} \, {\rm J} 
\approx 27.2 \, {\rm eV} \,.
\end{align}
\end{subequations}
Here, $|e|$ is the modulus of the elementary charge
(we reserve the symbol $e$ for the electron charge, see Ref.~\cite{JeLa2015epjd2}),
$\alpha$ is Sommerfeld's fine-structure constant,
while $m_e$ is the electron mass and $c$ 
denotes the speed of light. 
The fundamental atomic unit of energy is obtained 
by multiplying the fundamental atomic mass unit 
by the fundamental atomic unit of velocity, 
which is $\alpha \, c$. In atomic units, then, 
the reduced quantities fulfill the 
relations $c = 1/\alpha$ and $e = \hbar = m_e=1$, 
while $\epsilon_0 = 1/(4 \pi)$.

For completeness, we also indicate 
the explicit overall conversion from natural (n.u.) and atomic (a.u.) 
units to SI mksA for the polarizability, which reads as
\begin{align} 
\left. \alpha(\omega)] \right|_{\rm SI} = & \;
\frac{\epsilon_0 \, \hbar^3}{ m^3 \, c^3} \;
\left. \alpha(\omega)] \right|_{\rm n.u.}
\nonumber\\[0.133ex]
=& \; \frac{4 \pi \epsilon_0 \, \hbar^3}{\alpha^3 \, m^3 \, c^3} \;
\left. \alpha(\omega)] \right|_{\rm a.u.} \,.
\end{align} 
Judicious unit conversion helps to eliminate
conceivable sources of numerical error in the
final results for the friction coefficients.
The hydrogen and helium polarizabilities, in the 
natural and atomic unit systems, are well 
known~\cite{GaCo1970,Th1987,Pa1993,PaSa2000,MaSt2003,LaJeSz2004,Dr2005}.
From now on, for the remainder of the current 
section, we switch to atomic units.

In our numerical calculations, we 
concentrate on the evaluation of dielectric response function of
$\alpha$-quartz (SiO${}_2$), gold (Au), and calcium difluorite (CaF${}_2$). 
Indeed, a collection of references on optical properties of
solids has been given in 
Refs.~\cite{OrEtAl1983,OrEtAl1985,Pa1985,KaEtAl1962,DeEtAl1970,PSEtAl2009}. 
Following Ref.~\cite{LaDKJe2010pra}, we employ the following 
functional form for SiO$_2$ and CaF$_2$, which leads to a
satisfactory fit of the available data 
(see Tables~\ref{table1} and~\ref{table2}),
\begin{align}
\label{fit_formula}
\rho(\omega) =& \; \frac{\epsilon(\omega)-1}{\epsilon(\omega)+2}
= \frac{[n(\omega) + \ii \,  k(\omega)]^2-1}{[n(\omega)+\ii \, k(\omega)]^2+2}
\nonumber\\[0.133ex] 
\approx & \; \sum_{k=1}^n \alpha_k \,
\frac{\omega_k^2}{\omega_k^2 - \ii \, \gamma_k \, \omega-\omega^2} \,.
\end{align}
We have applied a model of this functional form to $\alpha$-quartz
(ordinary and extraordinary axis), Au and CaF${}_2$.
The form of $\rho$ is inspired by the 
Clausius--Mossotti equation, which suggests that the 
expression $[(\epsilon(\omega)-1)/(\epsilon(\omega)+2)]$
should be identified as a kind of polarizability 
function of the underlying medium. 
This function, in turn, exactly has the functional form 
indicated on the right-hand side of Eq.~\eqref{fit_formula}.
The dimensionless permittivity $\epsilon(\omega)$ is
obtained as $\epsilon(\omega) = (1 + 2 \, \rho)/(1 - \rho)$.
Also, it is useful to point out that 
the response function $(\epsilon(\omega)-1)/(\epsilon(\omega)+1)$,
whose imaginary part enters the integrand in 
Eq.~\eqref{eta1},
can be reproduced as follows,
\begin{equation}
\frac{\epsilon(\omega)-1}{\epsilon(\omega)+1} =
\frac{ 3 \rho(\omega) }{ \rho(\omega) + 2 } \,.
\end{equation}
Formula~\eqref{fit_formula} leads to a 
satisfactory representation of the
data for both infrared and ultraviolet absorption bands 
of SiO$_2$.

In order to model the dielectric response function of gold (Au),
we proceed in two steps. First, we employ a Drude model,
\begin{equation}
\label{drude}
\epsilon(\omega) = 1 - \frac{\omega_p^2}{\omega(\omega + \ii \gamma_p)} + 
\Delta \epsilon(\omega)
\end{equation}
with $\omega_p = 0.3330 \, E_h/h$ and $\gamma_p = 1.164 \times 10^{-3} \,
E_h/h$ (the specification in terms of $E_h/h$ is
equivalent to the use of atomic units).
For the remainder function $\Delta \epsilon(\omega)$, we find the 
following representation,
\begin{equation}
\label{Deltaeps}
\frac{\Delta \epsilon(\omega)-1}{\Delta \epsilon(\omega) + 2}
= \Delta \rho(\omega) \approx 1 - a 
+ \frac{a \, \omega_0^2}{\omega_0^2 - \ii \gamma_0 \omega - \omega^2}
\end{equation}
with $a = 1.5373$, $\omega_0 = 1.462 \, E_h/h$, and
$\gamma_0 = 4.550 \, E_h/h$.  In view of the asymptotics 
\begin{equation}
\Delta \rho(\omega) = 
1 + \frac{\ii \, a \, \gamma_0}{ \omega_0^2} \, \omega \,,
\qquad
\omega \to 0 \,,
\end{equation}
the functional form~\eqref{Deltaeps} ensures that 
the dielectric permittivity of gold,
as modeled by the leading Drude model term~\eqref{drude}, for $\omega\to 0$, 
retains its form of a leading term, equal to unity, 
plus an imaginary part which models the (nearly 
perfect) conductivity of gold for small driving frequencies.

Our discussion of atomic units provides us with an excellent opportunity to
discuss the natural unit of the normalized friction coefficient $\eta$.  In
order to convert $\eta$ from atomic to SI~mksA units, one needs to examine the
functional relationship $F_x = -\eta \, v_x$, where $v_x$ is the particle's
velocity.  The atomic unit of velocity is $\alpha \, c$, while the atomic unit
of force is equal to the force experienced by two elementary charges, which are
apart from each other by a Bohr radius. Denoting the atomic unit of force, for
which we have not found a commonly accepted symbol
in the literature, as $F_{\mbox{a.u.}}$, we have
\begin{equation}
F_{\mbox{a.u.}} = \frac{e^2}{4\pi\epsilon_0 \, a_0^2} = 
8.23872 \times 10^{-8} \, {\rm N} \,.
\end{equation}
The atomic unit $\eta_{\mbox{a.u.}}$
for the friction coefficient thus converts to SI~mksA units 
as follows,
\begin{equation}
\label{eta_au}
\eta_{\mbox{a.u.}} = 
\frac{F_{\mbox{a.u.}}}{\alpha \, c} = 
3.76594 \times 10^{-14} \; \frac{{\rm kg}}{{\rm s}} \,.
\end{equation}
For completeness, we also note the atomic units 
$\omega_{\mbox{a.u.}}$ and $\nu_{\mbox{a.u.}}$
of angular frequency and the cycles per second, 
respectively,
\begin{align}
\omega_{\mbox{a.u.}} = & \;
\frac{E_h}{\hbar} = 4.13414 \, 10^{16} \; \frac{{\rm rad}}{\rm s} \,,
\\[0.133ex]
\nu_{\mbox{a.u.}} = & \;
\frac{E_h}{h} = 6.57968 \, 10^{15} \; {\rm Hz} \,.
\end{align}

The data published in the reference volume
of Palik~\cite{Pa1985} for the optical
properties of solids relates to measurements at
room temperature. The integral~\eqref{eta1}
carries an explicit temperature dependence
in view of the Boltzmann factor, which appears
in disguised form (hyperbolic sine function in the
denominator), but there is also an implicit
temperature dependence of the dielectric
response function $[\epsilon(\omega)-1]/[\epsilon(\omega)+1]$,
which has been analyzed (for CaF${}_2$) in
Refs.~\cite{KaEtAl1962,DeEtAl1970,PSEtAl2009}.

For the SiO$_2$, gold and CaF$_2$ interactions investigated here,
we perform the calculations for temperatures around room temperature,
i.e., within the range $273 \, {\rm K} \leq T \leq 300 \, {\rm K}$.
We use the 
spectroscopic data from Tables~\ref{table1}---\ref{table2},
and employ the formula for the imaginary part of the 
polarizability given in
Eq.~\eqref{ImAlphaAU}, and the representation of the 
dielectric response function in Eq.~\eqref{fit_formula}.
Because of the narrow temperature range 
under study, this procedure is 
sufficient for $\alpha$-quartz and CaF$_2$.
For gold, we take into account the Drude model, as given 
in Eq.~\eqref{drude}. 
The uncertainty of our theoretical predictions 
should be estimated to be on the level of 10\% to 20\%,
in view of the necessarily somewhat incomplete
character of any global fit to discrete 
data on the dielectric constant and dielectric 
response function, which persists even if care is 
taken to harvest all available data from~\cite{Pa1985}.

{\em A priori}, the data in Palik's book~\cite{Pa1985} pertain to room 
temperature. For CaF$_2$, we may enhance the theoretical
treatment somewhat because the temperature dependence
of the dielectric response function
has been studied in 
Refs.~\cite{DeEtAl1970,OrEtAl1983,OrEtAl1985,PSEtAl2009}.
The dominant effect on the temperature dependence
of the dielectric response function of 
CaF$_2$ is due to the shift of the 
large-amplitude vibrational excitation at
$\omega_1 = 1.74 \times 10^{-3} \, {\rm a.u.}$ 
given in Table~\ref{table2}.
We find that the temperature-dependent data for the 
response function $[\epsilon(\omega) - 1]/[\epsilon(\omega) + 1]$
given in Fig.~10 of Ref.~\cite{PSEtAl2009}
can be fitted satisfactorily by introducing a 
single temperature-dependent
parameter in our fit function,
namely, a temperature-dependent width.
The replacement in terms of the parameters listed 
in Table~\ref{table2} is 
\begin{equation}
\gamma_1 \to \gamma_1 + a \, (T - T_0) \,,
\qquad
a = 4.97 \times 10^{-7} \frac{E_h}{h \, {\rm K}} \,,
\end{equation}
($4.97 \times 10^{-7}  \, {\rm a.u.}/{\rm K}$),
where $T_0 = 300 \, {\rm K}$ is the room-temperature reference 
point.

We finally 
obtain the friction coefficients given in Tables~\ref{table3}---\ref{table5}.
The normalized friction coefficient $\eta_0$ given in
Tables~\ref{table3}---\ref{table5} is indicated in atomic units, for a
distance of one Bohr radius from the surface. The $\calZ$ dependence and the
conversion to SI~mksA units is accomplished as follows: One takes the
respective entry for $\eta_0$ from Tables~\ref{table3}---\ref{table5}, 
multiplies 
it by the atomic unit of the friction coefficient given in 
Eq.~\eqref{eta_au} and corrects for the 
$1/\calZ^5$ and $1/\calZ^8$ dependences,
\begin{subequations}
\begin{align}
\label{eta_given}
\left. \eta^{(1)} \right|_{\rm SI} =& \; 
\left. \eta^{(1)}_0 \right|_{\rm a.u.} \, \left( \frac{a_0}{\calZ} \right)^5 \,
3.76594 \times 10^{-14} \; \frac{{\rm kg}}{{\rm s}} \,,
\\[0.133ex]
\left. \eta^{(2)} \right|_{\rm SI} =& \; 
\left. \eta^{(2)}_0 \right|_{\rm a.u.} \, \left( \frac{a_0}{\calZ} \right)^8 \,
3.76594 \times 10^{-14} \; \frac{{\rm kg}}{{\rm s}} \,.
\end{align}
\end{subequations}
This consideration should be supplemented by an example.
The backaction friction coefficients $\eta^{(2)}_x$ given in 
Tables~\ref{table3}---\ref{table5} are found to be 
numerically larger than the coefficients $\eta^{(1)}_x$
by several orders of magnitude, but they are 
suppressed, for larger atom-wall distances, by the 
functional form of the effect ($1/\calZ^8$ versus $1/\calZ^5$).
Let us consider the case of a helium atom
(mass $m_{\rm He} = 6.695 \times 10^{-27} \, {\rm kg})$,
at a distance 
\begin{equation}
\calZ_{20} = 20 \, a_0
\end{equation}
away from the $\alpha$-quartz surface
(extraordinary axis).
We employ the normalized friction coefficients
$\eta^{(1)}_0 = 8.81 \times 10^{-16}$ and
$\eta^{(2)}_0 = 4.80 \times 10^{-2}$ 
from Table~\ref{table3}, for a temperature $T = 298 \, {\rm K}$.
With 
\begin{equation}
u_0 = 3.76594 \times 10^{-14} \; {\rm kg} \, {\rm s}^{-1}
\end{equation}
being the atomic units of the friction coefficient,
the attenuation equation $F_x = -\eta \, v_x$ is solved by
\begin{subequations}
\begin{align}
\frac{\dd v_x}{\dd t} =& \; - \gamma \, v_x  \,,
\qquad
v_x(t) = v_x(0) \, \exp(-\gamma \, t) \,,
\\[0.133ex]
\gamma =& \; 
\left( \frac{\eta^{(1)}_{0x} \, u_0}{m_{\rm He}} 
\left( \frac{a_0}{\calZ_{20}} \right)^5 \right)
+
\left( \frac{\eta^{(2)}_{0x} \, u_0}{m_{\rm He}} 
\left( \frac{a_0}{\calZ_{20}} \right)^8  \right)
\nonumber\\[0.133ex]
=& \;
\left( 1.55 \times 10^{-9} \, {\rm s}^{-1} \right) +
\left( 10.55 \, {\rm s}^{-1} \right) 
\nonumber\\[0.133ex]
\approx & \; 10.55 \, {\rm s}^{-1} \,,
\end{align}
\end{subequations}
for ground-state helium atoms.
This corresponds to an attenuation time of 
$\tau = 0.0948 \, {\rm s}$,
in the functional relationship $\dd v_x/\dd t = v_x/\tau$.

%
%
\section{Conclusions}
\label{sec4}

In this paper, 
we have performed the analysis of the direct and backaction
friction coefficients in
Sec.~\ref{sec2}, to arrive at a unified formula for the quantum friction
coefficient of a neutral atom, in 
Eqs.~\eqref{eta1} and~\eqref{eta2}.
The numerical evaluation for the interactions 
of atomic hydrogen and helium with $\alpha$-quartz 
and calcium difluorite are described in
Sec.~\ref{sec3}.  The results in Tables~\ref{table3}---\ref{table5}
are indicated in
atomic units, i.e., in terms of the atomic unit of the friction coefficient,
which is equal to the atomic force unit (electrostatic force on two elementary
charges a Bohr radius apart), divided by the atomic unit of velocity [equal to
the speed of light multiplied by the fine-structure constant,
see Eq.~\eqref{eta_au}]. The conversion
of the entries given in Tables~\ref{table3}---\ref{table5} to SI units is
governed by Eq.~\eqref{eta_given}.
The friction coefficients indicated in Table~\ref{table4} for 
gold are smaller by several orders of magnitude than those 
for SiO$_2$ (Table~\ref{table3}) and CaF$_2$ (Table~\ref{table5}).

Finally, in Appendix~\ref{appa}, we illustrate the result on the basis of a
calculation of the Maxwell stress tensor, and verify that the zero-temperature
contribution to the quantum friction is suppressed in comparison to the main
term given in Eq.~\eqref{eta1}. In Appendix~\ref{appa}, we refer to the
zero-point/quantum fluctuations as opposed to the thermal fluctuations of the
electromagnetic field.

For a discussion of experimental possibilities to study the 
calculated effects discussed here, we refer to Ref.~\cite{JeLaDKPa2015}.
An alternative experimental possibility would
involve a laser interferometer~\cite{KeEkTuPr1991}. 
An interferometric apparatus has recently been proposed
for the study of gravitational interactions of anti-hydrogen atoms
(see Refs.~\cite{Ka2009AGE,AGELOI}); the tiny
gravitational shift of the interference pattern from atoms, after passing
through a grating, should enable a test of Einstein's equivalence principle for
anti-matter (this is the main conceptual idea of the AGE Collaboration, see
Ref.~\cite{AGELOI}).  Adapted to a conceivable quantum friction measurement, one might
envisage the installation of a hot single crystal in one arm of a laser atomic
beam interferometer, with a variable distance from the beam, in order to
measure the predicted ${\cal Z}^{-8}$ scaling of the effect.

%
%
\section*{Acknowledgments}

The authors acknowledge helpful conversations with 
G.~\L{}ach and Professor K.~Pachucki.
This research has been supported by the National Science Foundation 
Science Foundation (Grant PHY--1403973).
Early stages of this research have also been supported by 
a precision measurement grant from the
National Institute of Standards and Technology.

\appendix

%
%
\section{Quantum Friction for $\bm{T = 0}$}
\label{appa}

We start from the zero-temperature result for 
the quantum friction of two semi-infinite solids,
which is derived independently in Ref.~\cite{Pe1997}.
Indeed, from Eqs.~(15), (25) and~(54) of Ref.~\cite{Pe1997},
we have
\begin{align} 
\label{quantumfriction}
& F_x =\frac{\hbar\,S}{\pi^3}
\int_0^\infty\dd{k_\parallel}\,k_\parallel
\int_0^\infty\dd{k_\perp}\,e^{-2k\,\calZ}
\nonumber\\[0.133ex]
& \; \times \int\limits_0^{v_x \, k_\parallel}\dd \omega
\,{\rm Im}\left[ \frac{\epsilon_1(\omega)-1}{\epsilon_1(\omega)+1} \right]
\,{\rm Im}\left[ \frac{\epsilon_2(k_\parallel \, v_x -\omega)-1}
{\epsilon_2(k_\parallel \, v_x -\omega)+1} \right] \,.
\end{align} 
The quantum friction force for an atom can be 
obtained from the above formula by 
a matching procedure. 
Namely, for a dilute gas of atoms, which we assume to model the 
slab with subscript $1$, the relative permittivity can be written 
as follows,
\begin{equation}
\epsilon_1(\omega) = 1 + \frac{N_V}{\epsilon_0} \, \alpha(\omega) \,,
\end{equation}
where $\alpha(\omega)$ is the (dipole) polarizability,
and $N_V$ is the (volume) density of atoms.
Here, $\epsilon_1(\omega)$ is assumed to deviate from unity 
only slightly. We can then substitute
\begin{equation}
\frac{\epsilon_1(\omega)-1}{\epsilon_1(\omega)+1} \to
\frac{N_V}{2\epsilon_0}\alpha(\omega) \,.
\end{equation}
Here, $N_V = S^{-1} \, \dd N/\dd z$ is equal to 
the increase $\dd N$ in the number of atoms as we 
shift one of the plates by a distance $\dd z$ from the 
other. The factor $\dd N/\dd z$ can then be 
brought to the left-hand side where it reads as 
$F_\parallel(v) \dd z / \dd N$. Differentiating with respect
to $\dd z$, one obtains
$(\dd F_\parallel(v)/ \dd z) \, ( \dd z / \dd N) = 
\dd F_\parallel(v)/\dd N$, i.e.,
the force on the added atom.
The net result is that we have to 
differentiate $F_\parallel$ over $z$, 
and divide the result by $S \, N_V$, 
to obtain the force on the atom,
\begin{align}
& F_x =-\frac{\hbar}{\pi^3 \, \epsilon_0}
\int_0^\infty\dd{k_\parallel}\,k_\parallel
\int_{-\infty}^\infty\dd{k_\perp}\,k\, \ee^{-2k\,\calZ}
\nonumber\\[0.133ex]
& \qquad \times \int\limits_0^{v k_\parallel}\dd \omega
\; {\rm Im}\left[ \alpha(\omega) \right] 
\; {\rm Im}\left[ \frac{\epsilon(k_\parallel v_x -\omega)-1}%
{\epsilon(k_\parallel v_x - \omega)+1} \right] \,.
\end{align} 
In the limit of small velocities, i.e., $v_x \ll \calZ \, \omega_0$, 
where $\omega_0$ is the
first resonance frequency of either the atom $\alpha(\omega)$,
we can replace both the polarizability of the atom 
as well as the dielectric function of the solid by 
their limiting forms for small argument, i.e.,
small $\omega$  and small $\omega'=k_\parallel \, v_x - \omega$, 
can be replaced by
their low-frequency limits. We assume an atomic polarizability 
of the functional form 
\begin{equation}
\alpha(\omega) = 
\sum_n \frac{f_{n0}}{E_{n0}^2 - 
\ii \, \Gamma_n \, (\hbar \omega) - (\hbar \omega)^2} \,,
\end{equation}
where the oscillator strengths are denoted as $f_{n0}$ and
the $E_{n0}$ are the excitation frequencies of the atom.
For the zero-temperature quantum friction, the relevant limit is 
the limit of small angular frequency $\omega \ll E_{10}/\hbar$, and 
we assume that the first resonance dominates, with $\Gamma_1 \ll E_{10}$.
Under these assumptions, we can approximate
\begin{equation} 
\label{impolarizability}
{\rm Im}\left[ \alpha(\omega) \right] =
\sum_n \frac{f_{n0}}{E_{n0}^4} \, \Gamma_n \, \hbar\,\omega
\approx
\frac{\Gamma_1 \, (\hbar \omega)}{E_{10}^2} \, \alpha_0 \,.
\end{equation}
We have written $\alpha_0 = \alpha(0)$ for the static polarizability,
and we assume that the sum is dominated by the lowest 
resonance corresponding to the first excited state with $n=1$.
If the assumptions are not fulfilled, then the relationship 
\begin{equation} 
\alpha_0 = 
\frac{E_{10}^2}{\Gamma_1} \,
\sum_n \frac{f_{n0}}{E_{n0}^4} \, \Gamma_n 
\end{equation}
may serve as the definition of the quantity $\alpha_0$.
For the solid, we assume the functional form 
of a dielectric constant of a conductor,
which contains a term with zero resonance frequency in the 
decomposition of the dielectric function. We the have
(see also Ref.~\cite{JeLa2015epjd2}),
\begin{subequations}
\label{imepsiloncond}
\begin{align}
\epsilon(\omega) \sim & \;
1 - \frac{\omega_p^2}{\omega (\omega +\ii \gamma)} \,,
\\[0.133ex]
{\rm Im}\left[\frac{\epsilon(\omega)-1}{\epsilon(\omega)+1}\right]
\sim & \;
\frac{2\omega \, \gamma}{\omega_p^2}=
\frac{2\omega \, \epsilon_0}{\sigma_T(0)}\ ,
\end{align}
\end{subequations}
where $\sigma_T(0)$ is the temperature-dependent 
direct-current conductivity (for zero frequency).  
Substituting the results obtained in 
Eqs.~\eqref{impolarizability}) and~\eqref{imepsiloncond}) 
in Eq.~\eqref{quantumfriction} gives
\begin{align} 
F_x =& \;
-\frac{\hbar}{\pi^3\epsilon_0}
\frac{\Gamma_1 \,\alpha_0}{E_{10}^2}
\frac{2\gamma}{\omega_p^2}
\int_0^\infty\dd{k_\parallel}\,k_\parallel
\nonumber\\
& \; \times 
\int_{-\infty}^\infty\dd{k_\perp}\,k\, e^{-2k\,z}
\int_0^{v_x \, k_\parallel}\dd \omega\,\omega\,(k_\parallel v_x - \omega)
\nonumber\\
=& \; -\frac{45\hbar}{2^6\,\pi^2}
\frac{\Gamma_1\,\alpha_0\,\gamma}{\epsilon_0\,E_{10}^2\,\omega_p^2}
\frac{v_x^3}{z^7}
= -\frac{45\hbar}{2^6\,\pi^2}
\frac{\Gamma_1}{E_{10}^2} \, \frac{v_x^3}{\calZ^7} \,
\frac{\alpha_0}{\sigma_T(0)} \,,
\end{align}
with a $\calZ^{-7}$ dependence.
The $\epsilon_0$ factors cancel between the polarizability and the
conductivity. The result vanishes in the limit $\sigma_T(0) \to \infty$,
where many materials become superconducting [$\sigma(0) = \sigma_T(0) \to \infty$ for
$T \to 0$].

\end{document}